\documentclass[12pt]{article}
\usepackage{epsfig}
\title{\bf Vacuum replicas in two-dimensional QCD}
\author{P. J. A. Bicudo\thanks{E-mail: bicudo@ist.utl.pt}\hspace*{2mm}$^{\rm a}$,
A. V. Nefediev\thanks{E-mail: nefediev@heron.itep.ru}\hspace*{2mm}$^{\rm b}$}
\date{\small${\rm ^a}$ {\it Grupo Te\'orico de Altas Energias (GTAE),}\\
{\it Centro de F\'\i sica das Interac\c c\~oes Fundamentais (CFIF),}\\
{\it Departamento de F\'\i sica, Instituto Superior T\'ecnico,} \\
{\it Av. Rovisco Pais, P-1049-001 Lisboa, Portugal}\\
${\rm ^b}$ {\it Institute of Theoretical and Experimental Physics,}\\
{\it 117218, B.Cheremushkinskaya 25, Moscow, Russia}}
\newcommand{\be}{\begin{equation}}
\newcommand{\ee}{\end{equation}}

\newcommand{\too}{\mathop{\to}\limits_{N_C\to\infty}}
\newcommand{\vpint}{\int\makebox[0mm][r]{\bf --\hspace*{0.13cm}}}
\newcommand{\vpintt}{\int\makebox[0mm][r]{--\hspace*{0.08cm}}}
\newcommand{\ds}{\displaystyle}

\newcommand{\upp}[1]{\raisebox{1mm}{$#1$}}
\newcommand{\low}[1]{\raisebox{-1mm}{$#1$}}
\begin{document}
\maketitle

\begin{abstract}
Two-dimensional QCD is studied from the point of view of existence of
multiple chirally noninvariant solutions to the mass-gap equation. The
ground-state solution is reproduced and an infinite set of replica
solutions is discovered for this equation using the WKB quantisation procedure.
\medskip

{\noindent\scriptsize PACS: 12.38.Aw, 12.39.Ki, 12.39.Pn\\
Keywords: quark models, QCD, chiral symmetry breaking, Bogoliubov transformation}
\end{abstract}

In this paper, we revisit the two-dimensional model for QCD
\cite{tHooft,Einhorn} in the axial (Coulomb) gauge \cite{BG,ufn} and
address the question raised in a set of recent publications
\cite{replica1,replica2,replica3} as if replica solutions to the mass-gap
equation exist defining metastable chirally noninvariant phases of the
theory. For the four-dimensional potential model for QCD
\cite{Orsay,Lisbon} such solutions were discovered, for the
oscillator-type potential, in \cite{Orsay}, for the phenomenological
potential with the Coulomb interaction included, in \cite{replica1}, and,
finally, for an arbitrary power-like potential, in \cite{replica3}. A
similar conclusion was also made in a different approach \cite{oni}. In
this paper, we apply the analytical and numerical techniques, developed in
the work \cite{replica3} for the aforementioned four-dimensional model for
QCD, to a simpler theory --- the 't~Hooft model for QCD in two dimensions,
and demonstrate how replicas appear in this case. Thus we reproduce the
numerical solution for the BCS vacuum of the theory found in \cite{MingLi}
and build explicitly, as an example, the first \lq\lq excited" solution --- the replica
--- for the mass-gap equation. We argue that the number of such solutions
is infinite and evaluate their behaviour at the origin, considering the
mass-gap equation as a Shr{\" o}dinger-like equation and using the quasiclassical 
quantisation procedure for the latter.

Let us give necessary details of the two-dimensional model for QCD suggested by 't~Hooft in 1974
\cite{tHooft}. The theory is given by the Lagrangian
\be
L (x)= -\frac{1}{4}F^a_{\mu\nu}(x)F^a_{\mu\nu}(x) + \bar q(x)(i\gamma_{\mu}
\partial_{\mu} - gA^a_{\mu}t^a\gamma_{\mu} - m)q(x),
\label{lagrangian}
\ee
and the large-$N_C$ limit is understood, such that the effective coupling constant remains finite,
\be
\gamma=\frac{g^2N_C}{4\pi}\too {\rm const}.
\ee

It was demonstrated in \cite{Witten} that, in such a limit, the well-known no-go Coleman theorem
\cite{Coleman} can be bypassed, and the spontaneous breaking of the chiral symmetry (SBCS) may happen, provided the
limit $N_C\to\infty$ is taken prior to the chiral limit, $m\to 0$ \cite{Zhitnitskii}.

We gauge the theory by the condition
\be
A_1(x_0,x)=0,
\label{gauge}
\ee
so that the only nontrivial correlator of the gluonic fields, the gluon propagator, takes the form:
\be
D^{ab}_{00}(x_0-y_0,x-y)=-\frac{i}{2}\delta^{ab}|x-y|\delta(x_0-y_0),
\label{D}
\ee
yielding a linear confinement for the inter-quark interaction mediated by the two-dimensional
gluon. We use the following convention for $\gamma$ matrices:
\be
\gamma_0\equiv\beta=\sigma_3,~\gamma_1=i\sigma_2,~\gamma_5\equiv\alpha=\gamma_0\gamma_1=\sigma_1,
\ee
$\sigma$'s being the standard $2\times 2$ Pauli matrices. It was demonstrated in \cite{BG} that, in the axial
gauge (\ref{gauge}), the infrared divergences in the theory can be self-consistently regularised by the
principal value prescription, which is understood in all integrals below. If not stated otherwise, we consider
the chiral limit, $m=0$.

Following the standard technique, we build the Hamiltonian of the theory,
\be
H=-i\int dxq^{+}(t,x)\gamma_5\frac{\partial}{\partial x}q(t,x)-\frac{g^2}{4}\int
dxdy\;q^{+}(t,x)t^aq(t,x)\;|x-y|\;q^{+}(t,y)t^aq(t,y),
\label{H}
\ee
and arrange the normal ordering of the latter in terms of the dressed quark field,
\be
q_{\alpha i}(t,x)=\int\frac{dk}{2\pi}e^{ikx}[b_{\alpha}(k,t)u_i(k)+d_{\alpha}^\dagger (-k,t)v_i(-k)],
\label{qf}
\ee
\be
\ds\{b_{\alpha}(t,p)b^\dagger_{\beta}(t,q)\}=\ds\{d_{\alpha}(t,-p)d^\dagger_{\beta}(t,-q)\}=
2\pi\delta(p-q)\delta_{\alpha\beta},
\ee
\be
u(p)=T(p)\left(1 \atop 0 \right),\quad v(-p)=T(p)\left(0 \atop 1 \right),\quad
T(p)=\exp{\left[-\frac{1}{2}\theta_p\gamma_1\right]}.
\ee

The Bogoliubov (chiral) angle $\theta_p$ (shorthand notation for $\theta(p)$)
is odd and $\theta(p\to\pm\infty)\to\pm\pi/2$.
The normally ordered Hamiltonian (\ref{H}) splits into the vacuum energy, the quadratic, and quartic parts in
terms of the quark creation/annihilation operators,
\be
H=LN_C{\cal E}_v+:H_2:+:H_4:,
\label{Hh}
\ee
where $L$ is the one-dimensional volume; the vacuum energy density reads:
\be
{\cal E}_v =\int\frac{dp}{2\pi}Tr
\left\{\gamma_5p\Lambda_{-}(p)+\frac{\gamma}{4\pi}
\int\frac{dk}{(p-k)^2}\Lambda_{+}(k)\Lambda_{-}(p)\right\},\;
\Lambda_{\pm}(p)=T(p)\frac{1\pm\gamma_0}{2}T^\dagger(p).
\label{vac}
\ee
It is convenient to define the excess of the vacuum energy density over the free case
$\theta_{\rm free}(p)=\frac{\pi}{2}{\rm sign}(p)$,
\be
\Delta{\cal E}_v[\theta]={\cal E}_v[\theta]-{\cal E}_v[\theta_{\rm free}]=-\int\frac{dp}{2\pi}(p\sin\theta_p
-|p|)-\frac{\gamma}{4\pi}\int\frac{dpdk}{(p-k)^2}\cos[\theta_p-\theta_k].
\label{evac}
\ee

The actual form of the chiral angle is such that the quadratic part of the Hamiltonian (\ref{Hh}) becomes
diagonal in terms of the quark creation/annihilation operators, or alternatively, the vacuum energy takes
its minimal value:
\be
\frac{\delta}{\delta\theta_p}\Delta{\cal E}_v[\theta]=0,
\label{massgap}
\ee
which yields the mass-gap equation for the chiral angle $\theta_p$ \cite{BG},
\be
p\cos\theta_p=\frac{\gamma}{2}\vpint\frac{dk}{(p-k)^2}\sin[\theta_p-\theta_k],
\label{gap}
\ee
where, as mentioned above, the integral is defined by means of the principal value prescription.
For a given solution of Eq.~(\ref{gap}), the dressed quark dispersive law can be found from the system of
equations \cite{BG}:
\be
\left\{
\begin{array}{cc}
E_p\cos\theta_p=\frac{\upp{\ds \gamma}}{\low{\ds 2}}\ds\vpint\frac{dk}{(p-k)^2}\cos\theta_k\\[5mm]
E_p\sin\theta_p=p+\frac{\upp{\ds \gamma}}{\low{\ds 2}}\ds\vpint\frac{dk}{(p-k)^2}\sin\theta_k,
\end{array}
\right.
\label{system}
\ee
and reads
\be
E_p=p\sin\theta_p+\frac{\gamma}{2}\vpint\frac{dk}{(p-k)^2}\cos[\theta_p-\theta_k].
\label{E}
\ee

The mass-gap equation (\ref{gap}) is the main object of our studies. Following the method suggested in
\cite{replica3}, we rewrite (\ref{gap}) in the form of a Shr{\" o}dinger-like equation for the wave function
$\psi_p=\cos\theta_p$, in coordinate space, with the help of the relation (\ref{E}):
\be
\left[\hat{E}_p+\frac{\pi\gamma}{2} |x|\right]\psi_x=0,
\label{shrd}
\ee
where the following Fourier transformation was used:
\be
\vpint\frac{dk}{(p-k)^2}\psi_k=-\pi\int_{-\infty}^{\infty}dx|x|\psi_x e^{ipx}.
\ee

For further analysis it is convenient to introduce the effective mass of the quark $m_p$,
\be
\sin\theta_p=\frac{p}{\sqrt{p^2+m_p^2}},\quad\cos\theta_p=\frac{m_p}{\sqrt{p^2+m_p^2}},
\ee
and to rewrite the dispersive law $E_p$ in terms of the latter\footnote{Notice that Eqs.~(\ref{shrd}) and 
(\ref{Epp1}) follow immediately from the system (\ref{system}), form the first and the second equations,
respectively, and using the fact that, by definition of the principal value prescription, 
$\vpintt\frac{dk}{(p-k)^2}=0$.}:
\be
E_p=\sqrt{p^2+m_p^2}-\frac{\gamma}{2}\vpint\frac{dk}{(p-k)^2}\left(1-\frac{k\sqrt{p^2+m_p^2}}
{p\sqrt{k^2+m_k^2}}\right).
\label{Epp1}
\ee
so that
\be
E_p\approx |p|-\frac{\gamma}{|p|},
\label{Epp2}
\ee
where the approximation $m_p\ll p$ used in the last formula is valid for momenta $p$'s larger than some scale
$p_0$ generated by the function $m_p$, $p_0\sim m(p=0)$. Such a simplification is justified by the form of the
function $m_p$ which takes its maximal value at $p=0$ and then decreases fast for finite momenta. It is clear,
however, that in the approximate formula (\ref{Epp2}) we suppressed the positive contribution coming from the
first term on the r.h.s. of Eq.~(\ref{Epp1}) and enhanced the negative contribution coming from the second
term on the r.h.s. As a result, the approximate mass-gap equation takes the form:
\be
\left[\frac{\pi\gamma}{2} |\hat{x}|+|p|-\frac{\gamma}{|p|}\right]\psi_p=\varepsilon\psi_p,
\label{mglin1}
\ee
where we passed over to the momentum space, considering the quark
dispersive law $E_p$ as an effective potential. We are interested in the negative
eigenvalues $\varepsilon$. Besides that, according to the
definition of the wave function, $\psi_p=\cos\theta_p$, only even eigenstates
of the equation (\ref{mglin1}) should be selected.
We use the quasiclassical quantisation procedure to find the spectrum of
Eq.~(\ref{mglin1}),
\be
\frac{4}{\pi\gamma} \int_0^{p_{\rm max}}dp \left[\frac{\gamma}{p}-p+\varepsilon\right]=\pi\left(2n+\frac12\right),
\label{wkb}
\ee
and see that the integral on the l.h.s. diverges at $p=0$ since the approximation made above, $m_p\ll p$, 
fails for small momenta, where $m_p$ tends to a nonzero limit $m(p=0)\equiv\mu$. Therefore we introduce 
$\mu$ into the integral (\ref{wkb}), as a cut-off, for example, through the substitution
$|p|\to\sqrt{p^2+\mu^2}$ in the quark dispersive law. The resulting potential $V(p)=E_p$ appears sufficiently
attractive to possess bound states. Notice, however, that the cut-off $\mu$ itself is not a constant,
but it decreases fast for each next solution. Moreover, the solutions of the exact mass-gap equation 
(\ref{gap}) correspond to $\varepsilon_n=0$, for any $n$, so that one can reverse the arguments and use this 
condition to evaluate the behaviour of $\mu_n$ for an arbitrary $n$. The regularised quasiclassical 
integral, with $\varepsilon$ put to zero, is easily done, $I_{\rm WKB}\approx 
-\frac{4}{\pi}\ln(\mu_n/\sqrt{\gamma})+{\rm const}$, yielding
\be
\mu_n=\sqrt{\gamma}\exp{\left(-\frac{\pi^2}{2}n+\delta\right)},\quad n=0,1,2,\ldots,
\label{mun}
\ee
where $\delta$ is an $n$-independent finite correction to the leading regime. From Eq.~(\ref{mun}) one can easily estimate the
behaviour, at the origin, of the solution $\theta_p$ to the mass-gap equation (\ref{gap}). Indeed,
\be
\theta_p\mathop{\approx}\limits_{p\to 0}\sin\theta_p\approx\frac{p}{\mu_n}=
{\rm const}\frac{p}{\sqrt{\gamma}}e^{(\pi^2/2)n},
\label{teta}
\ee
where, in order to find the constant on the r.h.s., one needs either to find at least one solution to the mass-gap equation
numerically, or to estimate it using a more accurate regularisation procedure for the dispersive law and
evaluation of the quasiclassical integral. From the numerical solution for the ground-state vacuum found in
\cite{MingLi,replica1} this constant equals, approximately, 1.2. The accuracy of the approximate formula
(\ref{teta}) is expected to be better that 10\% even for the lowest states, as it
follows from the consideration in ref.~\cite{replica3}. 

\begin{figure}[t]
\epsfig{file=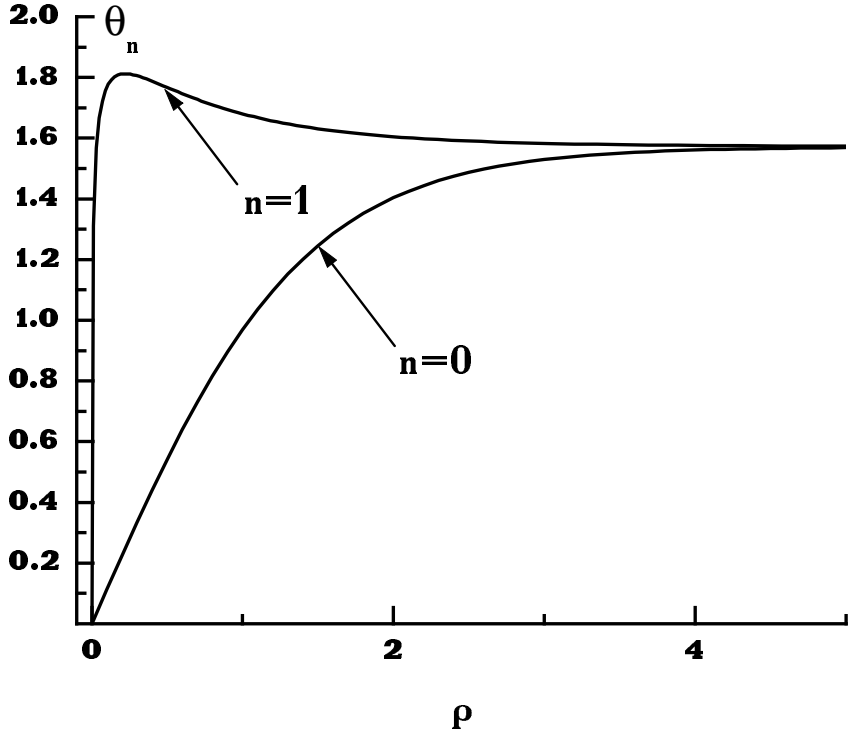,width=15cm}
\caption{The two lowest solutions to the mass-gap equation (\ref{gap}) corresponding to the BCS vacuum 
$(n=0)$ and to the first replica $(n=1)$. Since the chiral angle is an odd function, we 
plot it only for positive momenta. Momentum $p$ is given in the units of $\sqrt{\gamma}$.}
\end{figure}

In Fig.~1, as examples, we give the profiles of two
solutions to the mass-gap equation (\ref{gap}), found numerically, corresponding to the BCS vacuum 
$(\theta_0)$ and to the first replica $(\theta_1)$. The behaviour of these two solutions is in agreement with the
general formula (\ref{teta}).

From the formula (\ref{teta}) one can make several conclusions:
i) the solution (\ref{teta}) is purely nonperturbative since it involves $\sqrt{\gamma}$, whereas only integer
powers of $\gamma$ appear in any perturbative expansion;
ii) the mass-gap equation of the 't~Hooft model contains an infinite number of solution --- replicas ---
which appear due to a very peculiar sharp behaviour of the quark self-energy at small momenta, where the
second, negative, term on the r.h.s. in Eqs.~(\ref{Epp1}), (\ref{Epp2}) allows fast oscillations of the
chiral angle in the vicinity of the origin.
Each replica defines a phase of the theory with spontaneously broken chiral symmetry;
iii) the parameter defining SBCS, for example $\mu_n$, decreases very fast for each next solution,
$\mu_{n+1}/\mu_n\sim\exp{(-\pi^2/2)}\sim 10^{-2}$, which makes excited solutions extremely hard to find 
numerically;
iv) the series of replicas, for $n\to\infty$, converges to the
trivial chirally symmetric solution $\theta_{\rm trivial}(p)=\frac{\pi}{2}{\rm sign}(p)$, with a discontinuity at $p=0$, 
which can be
formally viewed as the limiting case $n=\infty$ of the formula (\ref{teta});
v) the vacuum energy density of replicas increase for each next solution converging to the trivial vacuum
energy. In contrast to four-dimensional QCD, where the trivial vacuum corresponds to zero vacuum energy, in two
dimensions the trivial vacuum energy density is infinite (see the discussion in \cite{replica1,ufn}).

As discussed in the papers \cite{replica1,replica2}, the existence of replicas is rather a general feature of
confining interactions, at least as far as confining potentials are concerned. In ref.~\cite{replica3} a 
similar
consideration was made for a generalised power-like potential in four dimensions and the same conclusion
was made concerning an
infinite number of solutions to the corresponding mass-gap equation, as a consequence of a very
peculiar behaviour of the quark self-energy in the infrared domain. Still it is instructive
to give an example of the model where there exists a unique solution to the mass-gap
equation. Using the same technique as in ref.~\cite{replica3}, one can easily
generalise the formula (\ref{teta}) for the case of an arbitrary power-like
potential in two-dimensions, $V(x)=K_0^{\alpha+1}|x|^\alpha$,
\be
\theta_p\mathop{\propto}\limits_{p\to 0}\frac{p}{K_0}\exp{\left[\frac{\pi n}{2}
\left(\frac{\Gamma\left(\frac12\right)\Gamma\left(\frac{2-\alpha}{2}
\right)}{\Gamma\left(\frac{1+\alpha}{2}\right)}\right)^{1/\alpha}\right]},
\label{teta2}
\ee
which reproduces Eq.~(\ref{teta}) for $\alpha=1$; $\Gamma(z)$ being the Euler
Gamma-function. One can see that, for
$\alpha\to 2$, that is, for the harmonic oscillator-type potential, the exponential factor
on the r.h.s. of Eq.~(\ref{teta2}) becomes infinite, so that only the case of $n=0$
--- the BCS vacuum --- survives. In order to illustrate the situation better let us apply the general method 
used in this paper to the harmonic oscillator-type potential directly. First, we derive the formula for the 
quark self-energy for such an interaction. Using Eq.~(\ref{E}), with the change 
$\gamma/(p-k)^2\to\pi K_0^3\delta''(p-k)$ (see \cite{replica1} for the details), one arrives at
\be
E_p=p\sin\theta_p-\frac{\pi}{2} K_0^3[\theta'_p]^2
\to |p|-\frac{\pi^3}{2} K_0^3[\delta(p)]^2
\to \sqrt{p^2+\mu^2}-\frac{\pi}{2}K_0^3 \frac{\mu^2}{(p^2+\mu^2)^2},
\label{E2}
\ee
where, in the last formula, we used the well-known representation for the delta-function, 
$\delta(p)=\lim_{\mu\to 0}\frac{1}{\pi}\frac{\mu}{p^2+\mu^2}$.
As a result, the mass-gap equation, similar to (\ref{mglin1}),
becomes a second-order Schr{\" o}dinger-like differential equation,
\be
\left[-\frac{\pi}{2} K_0^3\frac{d^2}{dp^2}+E_p\right]\psi_p=\varepsilon\psi_p,
\label{mgosc}
\ee
but the corresponding Bohr-Sommerfeld integral, for vanishing eigenenergy, monotonously
increase from zero to its maximal value of $\pi$, reached for $\mu\to 0$.
This makes a crucial difference between this theory and the 't~Hooft model, with linear confinement,
where the Bohr-Sommerfeld 
integral is not bounded from above and grows logarithmically with the decrease of the regulator $\mu$.
Therefore, for the two-dimensional linear confinement, the WKB quantisation condition can be satisfied for 
an arbitrary $n$, provided $\mu_n$ behaves properly (see Eq.~(\ref{teta}) above), whereas for the harmonic 
oscillator-type potential this condition can be fulfilled for the only value $n=0$. Thus we 
repeat the conclusion deduced from the general formula (\ref{teta2}). Indeed, the mass-gap
equation (\ref{mgosc}), which can be rewritten in the form
\be
p\cos\theta_p=-\frac{\pi K_0^3}{2}\theta''_p,
\ee
is known to have a unique nontrivial solution which can be
found either numerically \cite{replica1}, or it can be built approximately using the
method of separation of large- and small-momentum regions where the simplified equations are
solved analytically with the consequent matching of the solutions found in different
intervals (see \cite{Orsay} for the four-dimensional case). The result reads:
\be
\theta_p\approx\frac{\pi}{2}\left[1-3^{2/3}\Gamma\left(\frac23\right)Ai\left(\left(\frac{2}{\pi}\right)^{1/3}
\frac{|p|}{K_0}\right)\right]{\rm sign}(p),
\label{A4}
\ee
with $Ai(z)$ and $\Gamma(z)$ being the Airy and Gamma-function, respectively.
The formal reason for this behaviour of the theory can be easily deduced, as before, from the form of the quark 
self-energy (\ref{E2}), where the negative term has a delta-functional form and contributes only at $p=0$, 
whereas, in order to support replicas, it must give a sizable effect at finite momenta.

In four-dimensional QCD, the mass-gap equation for the harmonic oscillator-type potential is also 
differential, but it is known to support an infinite tower of replicas
\cite{Orsay}, similarly to other power-like potentials in four
dimensions \cite{replica3}. Extra singular terms in the quark self-energy appear, in this case, due to a
reacher structure of the interaction in four dimensions, namely, due to presence of
the angular variables. The formula similar to (\ref{teta2}) is derived in ref.~\cite{replica3}
and it contains an extra factor which cancels the singularity at $\alpha=2$ and
makes the corresponding limit smooth --- namely, this formula 
coincides with Eq.~(\ref{teta2}) with the only change $\Gamma\left(\frac{2-\alpha}{2}
\right)\to\Gamma\left(\frac{4-\alpha}{2}\right)$ \cite{replica3}, 
so that the general expression can be written with $\Gamma\left(\frac{D-\alpha}{2}\right)$, 
$D$ being the number of dimensions.
The case $\alpha=D$ is degenerate, that is, a unique solution to the mass-gap equation exists, 
but it can be reached only in two dimensions since, as argued in \cite{replica3}, for $\alpha>2$,
the mass-gap equation is divergent.

In contrast to four-dimensional QCD, where the limit $N_C\to\infty$ can be
relaxed and the effect of the QCD string breaking is expected to reduce the number of replicas making it
finite \cite{replica3}, in two dimensions, as mentioned above, the requirement $N_C=\infty$ is necessary to bypass the Coleman theorem and allow SBCS. 

Let us discuss a possible interpretation of replicas and their effect on two-dimensional observables. Notice that a whole tower of single-quark states can be built over each vacuum 
by a repeated application of the corresponding creation operators to the vacuum. Any such set provides a complete basis sufficient to saturate any multiquark state in the theory. An alternative, and in many cases more convenient, mathematical language for such sets is given by the well-known coherent states. Indeed, any nontrivial vacuum of the theory $|n\rangle$, $n=0,1,2\dots$, can be obtained from the trivial vacuum $|0\rangle_0$, corresponding to the chiral angle $\theta_p^{\rm triv}=\frac{\pi}{2}{\rm sign} (p)$, by means of the unitary operator $S_n$ \cite{Ls}:
\be
|n\rangle=S_n|0\rangle_0,\quad S_n=\exp{\left(Q^\dagger_n-Q_n\right)},\quad Q^\dagger_n=\frac{1}{2}\sum_\alpha\int \frac{dp}{2\pi}(\theta^n_p-\theta_p^{\rm triv})
b^\dagger_\alpha(p)d^\dagger_\alpha(-p),
\label{Sn}
\ee
or in the non-exponential form:
\be
|n\rangle=\prod_{p,\alpha}\left(\cos{\frac{\Delta\theta^n_p}{2}}+
C_{p\alpha}^\dagger\sin{\frac{\Delta\theta^n_p}{2}}\right)|0\rangle_0,\quad\Delta\theta^n_p=
\theta^n_p-\theta_p^{\rm triv},
\quad C_{p\alpha}^\dagger=b_\alpha^\dagger(p)d^\dagger_\alpha(-p).
\label{cs1}
\ee
It is easy to derive the vacuum orthogonality condition now, which reads
\be
\langle m|n\rangle={}_0\langle 0|S_m^\dagger S_n|0\rangle_0=\prod_{p,\alpha}\cos{\frac{\theta^n_p-\theta^m_p}{2}}=
\exp{\left[LN_C\int\frac{dp}{2\pi}\ln\left(\cos{\frac{\Delta\theta^{nm}_p}{2}}\right)\right]}\mathop{\to}\limits_{L\to\infty}0,
\label{ort1}
\ee
where $L$ is the one-dimensional volume. Thus any two vacua become orthogonal in the infinite volume.

Let us define coherent-like states $|\varphi,\beta\rangle$ \cite{replica2}:
\be
|\varphi,\beta\rangle=\prod_{p,\alpha}\left(\cos{\frac{\varphi_p}{2}+
C_{p\alpha}^\dagger[\beta_p]
\sin{\frac{\varphi_p}{2}}}\right)|0\rangle_0=\prod_{p,\alpha}|\varphi,\beta\rangle_{p\alpha},\quad
C_{p\alpha}^\dagger[\beta_p]\equiv e^{i\beta_p{\rm sign}(p)}C_{p\alpha}^\dagger,
\label{cs2}
\ee
which form an overcomplete set in the Fock space of the colourless quark-antiquark pairs, for any given momentum $p$ spanned by the two orthogonal states, $\{|0\rangle_0,C_{p\alpha}^\dagger[\beta_p]|0\rangle_0\}$. Therefore one can resolve the unitary operator in this space in terms of the states (\ref{cs2}):
\be
\hat{I}=\int[d\varphi][d\beta]|\varphi,\beta\rangle J\langle\varphi,\beta|\equiv\prod_{p,\alpha}\int_{-\pi}^{\pi}d\varphi_{p\alpha}\int_0^{2\pi}d\beta_{p\alpha}
|\varphi,\beta\rangle_{p\alpha} J_{p\alpha}(\varphi,\alpha){}_{p\alpha}\langle\varphi,\beta|.
\ee
where the choice of the Jacobean $J$ is not unique, the simplest form being just $2\pi^2$. The BCS vacuum state and all replicas belong to the set (\ref{cs2}) and correspond to a specific choice of the parameters $\varphi$ and $\beta$, $|n\rangle=|\Delta\theta_n,\frac{\pi}{2}\rangle$. Therefore the coherent-like states (\ref{cs2}) define an overcomplete basis, alternative to the complete set of multiquark states, each containing a fixed number of particles. Any vacuum replica can be chosen as the ground state to produce such an overcomplete basis, and all other replicas look like the coherent-like states in this basis. Notice that such a picture is maintained when the Hamiltonian of the theory is truncated and the $:H_4:$ part in the expression (\ref{Hh}) is neglected. Meanwhile as soon as the quartic part $:H_4:$ is included, the appropriate degrees of freedom of the theory, which minimise the vacuum energy and diagonalise the full Hamiltonian, are given by the quark-antiquark mesons. A second, generalised, Bogoliubov transformation is a possible way to arrive at the fully diagonalised Hamiltonian and at the bound-state equation, which plays the role similar to the mass-gap equation for the quadratic part of the Hamiltonian \cite{ufn}. The hadronic (mesonic) spectrum of the theory constructed over any replica vacuum contains tachyons, only the true BCS vacuum being free of the latter. Therefore the interpretation of replicas as stable vacua does not hold anymore, whereas, providing local minima of the vacuum energy, they may realise as metastable states in hadronic reactions \cite{replica1}. 
In what follows we confine our consideration of the trivial vacuum $|0\rangle_0$, so that for the sake of simplicity, we use the notation $\Delta\theta_p^n$ for the difference between the chiral angle for the $n$th replica $|n\rangle$ and the BCS vacuum $|0\rangle$.

In the paper \cite{replica2} a field theory approach to replicas in four dimensions is suggested which allows one to incorporate replicas, as local in space-time objects, into quark models. The given approach is easily adapted for the two-dimensional 't~Hooft model, so that the local operator creating the $n$th replica reads \cite{replica2}:
\be
S_n^T=T\exp\left[\int d^2x
f^{(n)}(x_0)\bar\psi(x_0,x)\left(-i\gamma_5\frac{d}{dx}\right)\psi(x_0,x)\right],
\label{Sop}
\ee
where the function $f^{(n)}(x_0)$ is even with the Fourier transform being such that
\be
f^{(n)}(2E_p)=\frac{\Delta\theta^n_p}{2p}.
\label{f}
\ee
On substituting the representation (\ref{qf}) for the quark field to Eq.~(\ref{Sop}), and with the help of the definition (\ref{f}), one easily recovers the form (\ref{Sn}) for the operator $S_n$.
 
From the operator (\ref{Sop}) one can easily derive the diagrammatic technique involving the new quark-quark-replica vertex, and the orthogonality condition (\ref{ort1}) can be easily reproduced then, via summation of an infinite set of fish-like diagrams, as a simple selfconsistency test of the approach. Performing a systematic expansion of all expressions in terms of the small difference $\Delta\theta^n_p$ and truncating the series at the second power of the latter, one can arrive at a simplified diagrammatic technique with an effective propagator of the replica:
\be
G_n(p_0,p)=\nu_n f^{(n)}(p_0)f^{(n)}(-p_0)(2\pi)\delta(p),
\label{Gn}
\ee
where $\nu_n=N_n/T$, $T$ being the full time of the process, $N_n$ giving the number of times the $n$th replica was excited during the time $T$. This propagator is attached to the quark lines through the two vertices $\gamma_5p$. An overall sum over $n$ is understood for the amplitude of the process in order to take into account the contribution of all replicas. In hadronic processes, the diagrams with replicas excitation provide corrections to standard amplitudes. This work is in progress now and will be reported elsewhere. 
The interested reader can find further details of the formalism in the paper \cite{replica2}.
To conclude, let us mention that with the replica solutions to the mass-gap equation found in this paper, the 't~Hooft model confirms its
status of a marvelous test-ground for the four-dimensional QCD where many interesting
phenomena can be observed and various approaches can be checked and tuned.

\bigskip
One of the authors (A.V.N.) would like to acknowledge the financial support of INTAS grants OPEN 2000-110 and
YSF 2002-49, as well as the grant NS-1774.2003.2, and the Federal Programme
of the Russian Ministry of Industry, Science and Technology No 40.052.1.1.1112.

\end{document}